\documentclass[12pt]{article}
\usepackage{graphicx}
\usepackage{amsmath}
\usepackage{amssymb}
\usepackage{hyperref}
\numberwithin{equation}{section}

\def\bC{\mathbb{C}}
\def\bZ{\mathbb{Z}}
\def\cN{\mathcal{N}}
\def\SU{\mathrm{SU}}
\def\SO{\mathrm{SO}}
\def\USp{\mathrm{USp}}

\def\diag{\mathop{\mathrm{diag}}}
\def\rank{\mathop{\mathrm{rank}}}
\def\tr{\mathrm{tr}}
\def\vev#1{\langle#1\rangle}

\newcommand{\beq}{\begin{equation}}
\newcommand{\eeq}{\end{equation}}
\newcommand{\beqa}{\begin{eqnarray}}
\newcommand{\eeqa}{\end{eqnarray}}
\newcommand{\CR}{\nonumber \\}

\begin{document}

\begin{titlepage}

\begin{flushright}
YITP-11-73\\
IPMU11-0130
\end{flushright}
\vskip 1.5cm
\begin{center}
{\Large \bfseries
Seiberg-Witten Geometries Revisited
}

\vskip 1.2cm

{\it --- Dedicated to the memory of Professor Sung-Kil Yang ---}

\vskip 1.8cm

Yuji Tachikawa$^1$ and Seiji Terashima$^2$

\bigskip\bigskip

$^1$ IPMU, University of Tokyo, \\
Chiba 277-8583 Japan

\bigskip

$^2$ Yukawa Institute for Theoretical Physics, Kyoto University, \\
Kyoto 606-8502, Japan
\vskip 1.5cm

\textbf{abstract}
\end{center}

We provide a uniform solution to 4d $\cN=2$  gauge theory with a single gauge group $G=A,D,E$ 
when the one-loop contribution to the beta function from any irreducible component $R$ of the hypermultiplets is less than or equal to half of that of the adjoint representation.
The solution is given by a non-compact Calabi-Yau geometry, whose defining equation is built from explicitly known polynomials $W_G$ and $X_R$,  associated respectively to the gauge group $G$ and each irreducible component $R$.
We provide many pieces of supporting evidence, for example by analyzing the system from the point of view of the 6d $\cN=(2,0)$ theory compactified on a sphere. 

\bigskip

\end{titlepage}

\setcounter{tocdepth}{2}
\tableofcontents

\section{Introduction}

It is by now well-known that the non-perturbative quantum effect of 4d
$\cN=2$ gauge theory can be captured by classical complex geometry
\cite{Seiberg:1994rs,Seiberg:1994aj}.
This geometry is usually called  the Seiberg-Witten curve when it is a curve, or 
just the Seiberg-Witten geometry when it is not.
Many methods to obtain the solutions  were devised over the years:
\begin{itemize}
\item  One is to consider the geometry as the spectral
curve of the integrable system
\cite{Gorsky:1995zq,Martinec:1995by,Donagi:1995cf,D'Hoker:1999af}. 
It is, however, not straightforward to pin down the integrable system given the gauge theory. 
For reviews on this approach, see e.g.~\cite{Itoyama:1995nv,Donagi:1997sr,D'Hoker:1999ft}
\item Another is to use the Calabi-Yau compactification of the string theory. Originally, the solutions were extracted from the decoupling limit of compact Calabi-Yaus \cite{Kachru:1995wm,Kachru:1995fv,Billo:1997vt}, but eventually the method was distilled into the geometric engineering \cite{Klemm:1996bj,Katz:1996fh}, which directly gives a non-compact Calabi-Yau which is fibered by ALE spaces, or equivalently 6d $\cN=(2,0)$ theory put on a Riemann surface. This requires the identification of the heterotic dual to the Type IIB on non-compact Calabi-Yau manifolds, and it is not always easy. For a review, see e.g.~\cite{Lerche:1996xu}.
\item Yet another is to use branes in the type IIA theory which is then lifted to M-theory \cite{Witten:1997sc}. 
This method suits very well classical gauge groups with usual representations, but it is not very helpful for exceptional cases. For a review on this approach, see e.g.~\cite{Giveon:1998sr}.
\item  Lately, it was realized that  6d $\cN=(2,0)$ compactified on a Riemann surface can be understood from the properties of codimension-2 defects of the 6d theory \cite{Gaiotto:2009we,Gaiotto:2009hg}. Then, finding a Seiberg-Witten solution reduces to the identification of the combination of the codimension-2 defects.
\end{itemize}
These techniques are all inter-related, and each is complementary to another. 
The state of the art is that, for almost all  of the choice of the gauge group and the matter content, at least one of the method is applicable, and the Seiberg-Witten solution can be found, and indeed the solution to most of the choices has been written down. 

Glancing through the solutions available in the literature \cite{Klemm:1994qs,Argyres:1994xh,Hanany:1995na,Argyres:1995wt,Brandhuber:1995zp,Martinec:1995by,Argyres:1995fw,Hanany:1995fu,Brodie:1997qg,Aganagic:1997wp,Terashima:1998iz,Terashima:1998fx,Hashiba:1999ss,Argyres:2002xc}, it was noted in \cite{Terashima:1998fx} that the solution to a theory with single simply-laced gauge group $G=A,D,E$ almost always is given  uniformly  by the Seiberg-Witten geometry \begin{equation}
z + \frac{\Lambda^{2h^\vee}}z \prod_R \Lambda^{-b_R} X_R(x_1,x_2,x_3;w_i;m)  = W_G(x_1,x_2,x_3;w_i).\label{SWgeometry}
\end{equation}  where
\begin{itemize}
\item $h^\vee$ is the dual Coxeter number of $G$,
\item $W_G$ is the equation of the ALE space of type $G$ deformed by the Casimirs $w_i$ given in Table~\ref{gaugedata}, \ref{gaugeeq}.
\item $b_R$ is the contribution to the one-loop beta function from the hypermultiplet in the irreducible representation $R$, given in Table~\ref{repdata},
\item and $X_R$ is the polynomial representing the hypermultiplet in the irreducible representation $R$, tabulated in the Appendix \ref{XR}.
\end{itemize}
Note that we can consider half-hypermultiplets when $R$ is a pseudo-real representation. We will denote them by $\frac12R$, and $b_{\frac12R}=\frac12 b_R$.

$X_R$ is known for all $R$ for which $b_R\le h^\vee$, and there is a
 reason to suspect that a matter representation $R$ with $h^\vee<b_R\le
 2h^\vee $, although still
asymptotically free or conformal,
is not realizable in this form. 
We call the former `nice' representations $(b_R\le h^\vee)$, and the latter `reasonable' representations. $(h^\vee <b_R \le 2h^\vee)$. 
The main objective of this paper is to give a rationale behind this regularity of the solution to the $\cN=2$ gauge theory  with single simply-laced gauge group and  nice hypermultiplets.
Here we note that most of the possible matter representations are 
in the ``nice'' representations.
Indeed, as seen from Table~\ref{repdata}, 
the only non-nice reasonable representations are 
only the 2-index symmetric tensor for $\SU(n)$,
the 3-index antisymmetric tensor for $\SU(7)$ and $\SU(8)$,
the spinor for $\SO(14)$,
and the adjoint representations.\footnote{
The Seiberg-Witten curves which include the 2-index symmetric tensor
and the adjoint representations are given in \cite{Landsteiner:1997ei} and \cite{D'Hoker:1998yi}, 
respectively. Thus, the exact solutions for any matter
content for the gauge group are known except for the ones with the spinor
representation of $\SO(14)$ and the 3-index antisymmetric tensor of $\SU(7)$ and $\SU(8)$.
We comment on this point again in Sec.~\ref{conclusions}.}

\begin{table}
\[
\begin{array}{rc|c|c|c|c|l}
&G&h^\vee & x_1 & x_2 & x_3 & w_i \\
\hline\hline
\SU(n)= &A_{n-1} & n & 1 &  n/2 & n/2  & 2,3,\ldots,n \\ 
\SU(n)=&A_{n-1}' & n & 1 &  n-2 & 2  & 2,3,\ldots,n\\
\SO(2n)=&D_n & 2n-2 & 2 & n-2 & n-1 & 2,4,\ldots,2n-2 ; n \\
&E_6 & 12 & 3 & 4 & 6 &2,5,6,8,9,12 \\
&E_7 & 18 & 4 & 6 & 9 &2,6,8,10,12,14,18 \\
&E_8 & 30 & 6 & 10 & 15 & 2,8,12,14,18,20,24,30 
\end{array}
\]
\caption{The dual Coxeter number for the gauge group $A_{n-1}$, $D_n$ and $E_{6,7,8}$, together with the mass dimension of $x_i$ in the corresponding ALE spaces and of the Casimirs $w_i$. \label{gaugedata}}
\end{table}

\begin{table}
\begin{align*}
W_{A_{n-1}}& =x_1^n+x_2^2+x_3^2  +  w_2 x_1^{n-2}+w_3 x_1^{n-3} + \cdots +w_n, \\
W_{A_{n-1}'} &= x_1^n+x_2 x_3  + w_2 x_1^{n-2}+w_3 x_1^{n-3} + \cdots +w_n, \\
W_{D_{n}}  &= x_1^{n-1} + x_1 x_2^2 - x_3^2  + w_2 x_1^{n-2}+ w_4 x_1^{n-3} + \cdots+ w_{2n-2} + \tilde w_n x_2, \\
W_{E_6}&= x_1^4+ x_2^3+x_3^2+ w_2 x_1^2 x_2 + w_5 x_1 x_2 + w_6 x_1^2 + w_8 x_2 + w_9 x_1 + w_{12},\\
W_{E_7}&=  x_1^3{x_2} +x_2^3+x_3^2 \nonumber\\
&\qquad + w_2 {x_2}^2 {x_1} + w_6 {x_2}^2  + w_8 {x_2} {x_1} + w_{10} {x_1}^2 + w_{12} {x_2} + w_{14} {x_1}+ w_{18} \\
W_{E_8} &= x_1^5 + x_2^ 3 + x_3^2  +  \nonumber \\
& \qquad + w_2 x_2 x_1^3 + w_8 x_2 x_1^2 + w_{12} x_1^3 +w_{14} x_2 x_1 + w_{18} x_1^2 + w_{20} x_2 + w_{24} x_1 + w_{30}
\end{align*} 
\caption{
The equation defining the ALE spaces. The explicit formula for $w_i$ via the Cartan of $G$ was given in \cite{KatzMorrison,Shioda,Noguchi:1999xq}, and is reproduced in Appendix~\ref{XR}.
 \label{gaugeeq} }
\end{table}

Although $X_R$ for all nice $R$ can be found and will be tabulated in this paper,  the mathematics behind them is not yet as clear as that for $W_G$. One way to obtain a handle to $X_R$ is to view it from the perspective of punctures of 6d $\cN=(2,0)$ theory, which is most straightforward when $\sum_{R_i} b_{R_i}=h^\vee$ \cite{Gaiotto:2009we,Chacaltana:2010ks,Chacaltana:2011ze}. In this case we expect that $\cN=(2,0)$ theory compactified on a sphere with three regular punctures realize free hypermultiplets in the representation $\oplus R_i$. 
We give a detailed analysis of two cases, one involving the three-index antisymmetric tensor of $\SU(6)$ and another involving $\mathbf{56}$ of $E_7$.

The rest of the paper is organized as follows: in Sec.~\ref{general}, we interpret the polynomial $X_R$ as giving a Calabi-Yau which represents a free hypermultiplet in the representation $R$, and give an explanation of the uniformity of the solution \eqref{SWgeometry}. We also discuss how $X_R$ for different $G$ and $R$ is related to each other. In Sec.~\ref{comparison}, we perform three detailed case studies: the first is the 2-index antisymmetric tensor of $\SU(N)$, 
the second is $\mathbf{20}$ of $\SU(6)$, and the third is $\mathbf{56}$ of $E_7$.
Among others, we will find that the same $X_R$ can arise from a completely different combination of punctures from the point of view of the 6d $\cN=(2,0)$ theory. We conclude with a short discussion in Sec.~\ref{conclusions}. The Appendix \ref{XR} contains the list of all $X_R$.
In Appendix B \ref{cg} we describe the details of obtaining the curve from the geometry.

\section{Generalities}\label{general}
\subsection{Motivating examples}
\subsubsection{$\cN=2$ gauge theory without matters}
Let us start by the analysis of the solution to the pure $\cN=2$
gauge theory with gauge group $G=A,D,E$ \cite{Klemm:1994qs,Argyres:1994xh,Brandhuber:1995zp,Martinec:1995by,Klemm:1996bj,Lerche:1996an} : \begin{equation}
\Lambda^{h^\vee} z + \frac{\Lambda^{h^\vee}}z = W_G(x_1,x_2,x_3;w_k) \label{pure}
												 \end{equation} where the Seiberg-Witten three-form is given by \begin{equation}
\omega=\frac{1}{2\pi i}\oint \frac{dx_1\wedge dx_2\wedge dx_3}{W_G (x_1,x_2,x_3;w_k) -\Lambda^{h^\vee}(z+1/z) } \wedge \frac{dz}z
\end{equation} 
where the contour integral is taken around the poles of the denominator.

It is helpful to recall the property of the deformed ALE space $W_G=0$. It is quasihomogeneous in the variables $x_i$ and the coefficients $w_k$ whose degrees are given in Table~\ref{gaugedata};
we use the convention that $w_k$ has degree $k$.  When $w_i$ are generic, there are $r=\rank G$ two-cycles $C_1,\ldots C_r$ intersecting according to the Dynkin diagram of $G$. Then, we can define an element $\phi$ in the Cartan of the Lie algebra of $G$ such that \begin{equation}
\alpha_i\cdot \phi = \int_{C_i} \oint \frac{dx_1\wedge dx_2\wedge dx_3}{W_G}
\end{equation} where $\alpha_i$ is the $i$-th simple root. 
Then $w_k$ is given by the degree-$k$ Casimir constructed from $\phi$ \cite{Shioda,KatzMorrison}.
The geometry $W_G=0$ becomes singular when $\alpha\cdot\phi=0$ for a root $\alpha$. 
It is well known that the low-energy limit of the Type IIB string on $W_G=0$ gives 6d $\cN=(2,0)$ theory of type $G$.

From this viewpoint, the geometry for the pure gauge theory \eqref{pure} can also be written as \begin{equation}
W_G(x_1,x_2,x_3;w_k(z))=0 \qquad \text{where}\ \Bigl\{ 
\begin{array}{r@{\,}l}
 w_k(z)&=w_k\hfill (k<h^\vee), \\
 w_{h^\vee}(z)&=-\Lambda^{h^\vee} z +w_{h^\vee} -\Lambda^{h^\vee}/z.
\end{array}
\end{equation} 
This describes how the ALE space is deformed as one changes $z$.
It  can be regarded as the compactification of the 6d $\cN=(2,0)$ theory of type $G$ on the cylinder parameterized by $z$.
Then $w_k(z)$ is the worldvolume fields of the 6d theory which depends on the position on the cylinder.

\begin{figure}
\[
\begin{array}{cr}
a) & \vcenter{\hbox{\includegraphics[scale=.6]{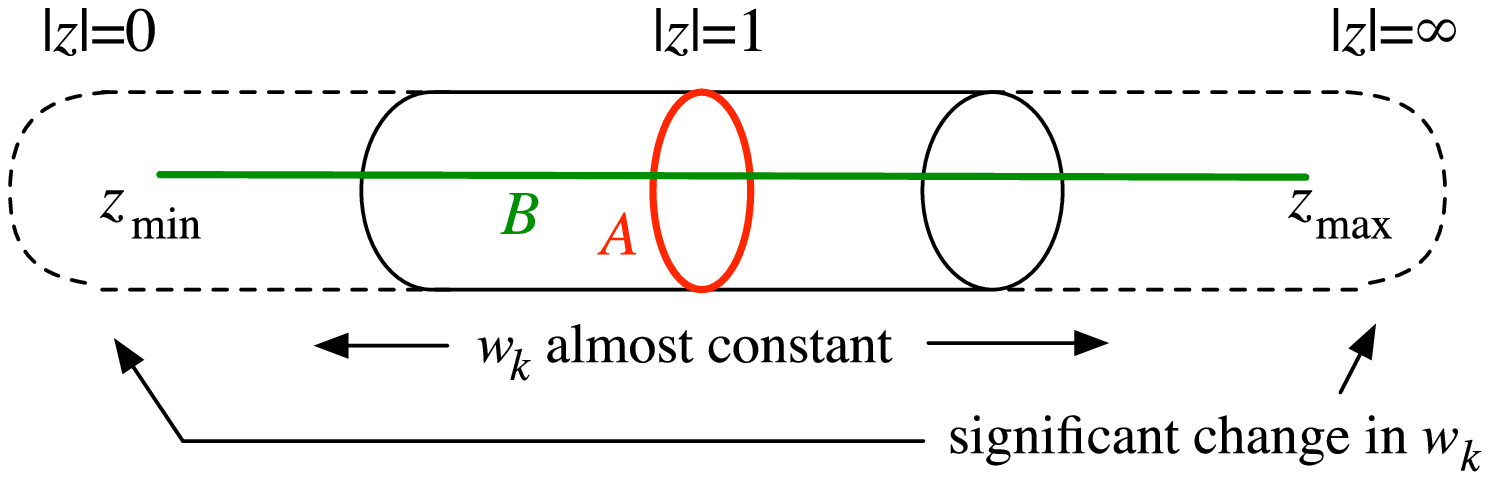}}} \\
b) &\vcenter{\hbox{\includegraphics[scale=.6]{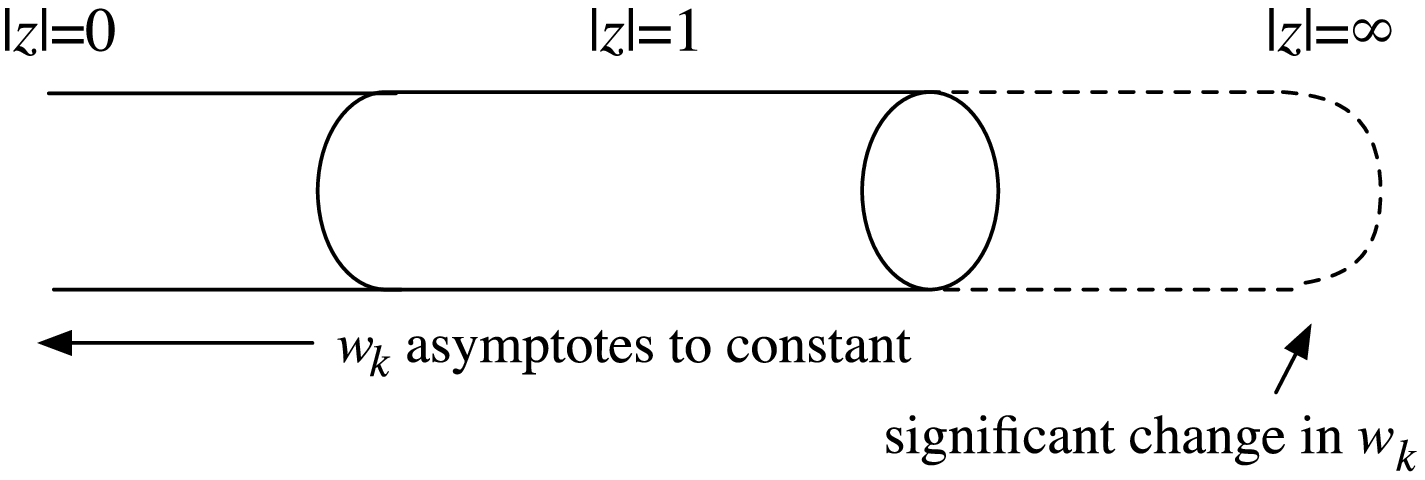}}} 
\end{array}
\]
\caption{a) The base of the Seiberg-Witten geometry in the weakly-coupled region. b) a half of the Seiberg-Witten geometry. This can be though of as the geometry representing free hypermultiplets. \label{geometry}}
\end{figure}

Let us choose the vev so that $w_k\sim a^k$. When the system is weakly coupled, i.e.~$\Lambda \ll a$, the geometry can be visualized as in Fig.~\ref{geometry}a).  When $|z|\sim O(1)$, $w_k(z)$ is almost constant. Let us take an element $\phi$ in the Cartan of the Lie algebra of $G$ so that $w_k=w_k(\phi)$.
Take a three-cycle of the form $A_i=A \times C_i$ where $A$ is a cycle wrapping around the cylinder once.
Then \begin{equation}
\int_{A_i} \omega \sim \alpha_i\cdot \phi.\label{W}
\end{equation}
This means that a D3-brane wrapped around $A_i$ gives rise to the W-boson of the gauge group $G$ corresponding to the root $\alpha_i$ \cite{Klemm:1996bj,Hollowood:1997pp,Gaiotto:2009hg}. 

The geometry is terminated by the  divergence of $w_{h^\vee}(z)$.
Take the path $B$ connecting $z=z_\text{max}$ and $z=z_\text{min}$ where the  two-cycle $C_i$ shrinks.
The geometry is significantly modified when the right hand side on \eqref{pure} is of the same order with the left hand side. So we can estimate $|z_\text{max}|\sim |1/z_\text{min}|\sim (a/\Lambda)^{h^\vee}$.

Consider the three-cycle of the form $B_i=B\times C_i$  see Fig.~\ref{geometry}a). Then we find \begin{equation}
\int_{B_i} \omega \sim (\alpha_i\cdot \phi) \frac{1}{2\pi i}\int_{z_\text{min}}^{z_\text{max}}\frac{dz}z\sim  \frac{2h^\vee}{2\pi i}(\alpha_i\cdot\phi)\log \frac{\Lambda}{a} \label{monopole}
\end{equation} which is the mass of the BPS 't Hooft-Polyakov monopole. The ratio of the masses \eqref{W} and \eqref{monopole}, \begin{equation}
\tau(a) = \frac{2h^\vee}{2\pi i} \log\frac\Lambda a
\end{equation} correctly gives the one-loop running of the low-energy gauge coupling on the Coulomb branch.

The geometry around $|z|\sim \infty$ can be studied by dropping the term inversely linear in $z$ in \eqref{pure}: \begin{equation}
\Lambda^{h^\vee} z = W_G(x_1,x_2,x_3;w_i),
\end{equation} see Fig.~\ref{geometry}b). This should represent the
presence of no vector multiplet at all.
Indeed this geometry is smooth for all values of $w_i$, because of the linear term in $z$. 

\subsubsection{$\SU(n)$ with fundamental hypermultiplets}

Let us next consider the geometry for the $\SU(n)$ gauge theory with $n_f<2n$ fundamental flavors with masses $m_i$ \cite{Hanany:1995na,Argyres:1995wt}: \begin{equation}
z+\frac{\Lambda^{2n-n_f}}z \prod_{i=1}^{n_f} (x_1-m_i) = x_2^2+x_3^2+ P(x_1)
\end{equation} where $P(x)=\prod (x-a_i) = x_1^n+w_2 x_1^{n-2} + \cdots+ w_n$.
We can split $n_f=n'_f+n''_f$ so that $n'_f,n''_f\le n$. Then we can write the same geometry as \begin{equation}
\Lambda^{n-n''_f}z \prod_{i=1}^{n''_f} (x_1-m_{i}) +\frac{\Lambda^{n-n'_f}}z \prod_{i=1}^{n'_f} (x_1-\tilde m_i) = x_2^2+x_3^2+ P(x_1).\label{foo}
\end{equation} where we redefined $z$.
Now the degree of the left hand side is less or equal to that of the right hand side, and thus the system can be thought of as the compactification of 6d $\cN=(2,0)$ theory of type $A_{n-1}$. When $\Lambda$ is very small, $w_k(z)$ is again almost constant around $|z|\sim O(1)$, which again fits the situation shown in Fig.~\ref{geometry}, a). 
We can find the three-cycles representing the W-bosons and the monopoles as before; the only difference is that now $z_\text{max}\sim (a/\Lambda)^{n-n_f''}$ 
and $z_\text{min}\sim (\Lambda/a)^{n-n_f'}$. Then the ratio of the masses is now \begin{equation}
\tau(a) = \frac{2n-n_f'-n_f''}{2\pi i} \log\frac\Lambda a,
\end{equation} correctly reproducing the one-loop running.

The divergences of $w_k(z)$ at $|z|\sim 0$ and $|z|\sim \infty$ should then represent the presence of $n'_f$ and $n''_f$ hypermultiplets in respective regions.
To isolate the behavior at $|z|\sim \infty$, we drop the term inversely proportional to $z$ in \eqref{foo}, and consider the geometry \begin{equation}
\Lambda^{n-n_f''} z \prod_{i=1}^{n_f''} (x_1-m_{i}) = x_2^2+x_3^2+ \prod_{i=1}^n (x_1-a_i) .
\end{equation} 
The geometry can be visualized as in Fig.~\ref{geometry}b).
As can be easily checked, this geometry becomes singular when $a_i=m_j$ for some $i$ and $j$, signifying the presence of the fundamental hypermultiplets with masses $m_i$. Note also that the prefactor $\Lambda^{n-n_f''}$ was responsible for reproducing the one-loop running. 
This geometry will be obtained by the weak coupling limit, i.e. 
$\Lambda \rightarrow 0$ although only the $n''_f$ hypermultiplets can be represented. 

\begin{table}
`Nice' representations: $b\le h^\vee$ 
\[
\begin{array}{rc|crc|c|c}
& G& & &R   & b & X_R \\
\hline \hline
\SU(n)=&A_{n-1}& &\text{fund.} &\mathbf{n}& 1  & \eqref{fund}\\
&& &\text{2-index antisym.} &\mathbf{\frac{n(n-1)}2}& n-2 & \eqref{asym}\\
\SU(6)=& A_5 & \text{half}& \text{3-index antisym.} &\mathbf{20} & 3 &\eqref{20}\\
\SO(n)=&D_n & & \text{vector} &\mathbf{2n}& 2 &\eqref{vector}\\
\SO(8)=&D_4 & & \text{spinor} &\mathbf{8s} & 2 &\eqref{8s}\\
\SO(8)=&D_4 & & \text{conj. spinor} &\mathbf{8c} & 2 &\eqref{8c}\\
\SO(10)=&D_5 & & \text{spinor}& \mathbf{16} & 4 &\eqref{16}\\
\SO(12)=&D_6 & \text{half} & \text{spinor} &\mathbf{32s} & 4 &\eqref{32}\\
\SO(12)=&D_6 & \text{half} & \text{conj. spinor} &\mathbf{32c} & 4& \eqref{32c}\\
&E_6 &  && \mathbf{27} & 6 &\eqref{27}\\
&E_7 & \text{half} & &\mathbf{56} & 6 &\eqref{56}\\
\end{array}
\]
`Reasonable' representations: $h^\vee < b \le 2 h^\vee$ 
\[
\begin{array}{rc|c|c}
& G& R   & b \\
\hline \hline
\SU(n)=&A_{n-1} 
& \text{2-index sym.} & n+2 \\
&& \text{adj.} & 2n \\
\SU(7)=& A_6 & \text{3-index antisym.} &  10\\
\SU(8)=& A_6 & \text{3-index antisym.} &  15\\
\SO(2n)=&D_n & \text{adj.} & 4n-4 \\
\SO(14)=&D_7&\text{spinor} & 16 \\
&E_6 & \text{adj.} & 24 \\
&E_7 & \text{adj.} & 36 \\
&E_8 & \text{adj.} & 60 \\
\end{array}
\]
\caption{List of `nice' and `reasonable' representations of $G=A_{n-1}$, $D_n$ and $E_{6,7,8}$ \label{repdata}}
\end{table}

\subsection{Hypermultiplet geometries}
The discussion so far suggests the following general picture: for a hypermultiplet in representation $R$ of $G$ with mass $m$, we should find a Calabi-Yau geometry\footnote{ Calabi-Yau geometries which produce massless hypermultiplets in various representations when used in Type IIA compactifications were identified in \cite{Katz:1996xe,Morrison:2011mb}.  In our case, the Calabi-Yau geometries should be used in Type  IIB compactifications. These should be  mirror to each other, but since they are non-compact and typically non-toric, it is hard to show that it is indeed the case. }  representing it of the form \begin{equation}
\Lambda^{h^\vee-b_R} zX_R(x_1,x_2,x_3;w_k;m) = W_G(x_1,x_2,x_3;w_k)
\end{equation} such that there is a three-cycle $C_v$  for each weight $v$ of $R$ such that \begin{equation}
\int_{C_v} \omega = v\cdot \phi-m
\end{equation} where $\phi$ is such that $w_k=w_k(\phi)$. Finding three-cycles is a big problem, so in practice we impose a weaker condition, that
the geometry becomes singular whenever  \begin{equation}
v\cdot \phi = m 
\end{equation} for a weight $v$ of $R$. It is not, a priori, obvious that we can find such a geometry for all $R$, or why it should be linear in $z$. 

First, in order for the left hand side to be a deformation of the
singularity, $b_R$ can be at most $h^\vee$, or in our terminology, $R$ needs to be `nice'.  
Therefore, the adjoint
representation for which $b=2h^\vee$ is out of the question in this
approach. Fortunately, one finds that most of the representations which can be used to
 construct asymptotically free or conformal theory with simply-laced gauge group $G=A,D,E$ are nice.
The  exceptions are the adjoint representation of each $G$, the symmetric 2-index tensor representation of $\SU(N)$, the antisymmetric 3-index representations of $\SU(7)$ and $\SU(8)$, and the spinor representation of $\SO(14)$.  Therefore, representations with $b_R\le h^\vee$ are not that a big restriction, see Table~\ref{repdata}.

Second, the table shows that there are two types of irreducible representations with $b_R\le h^\vee$: one which comes in infinite series, fundamental and two-index antisymmetric tensor of $\SU(n)$, and  vector of $\SO(2n)$.  
There are only finite number of exceptions, which we call \emph{exceptional representations}.
The curves for the infinite series are standard \cite{Hanany:1995na,Argyres:1995wt,Argyres:1995fw,Hanany:1995fu,Argyres:2002xc}, and can be translated into corresponding polynomials $X_R$. 
As for the exceptional representations, the biggest  is $\mathbf{56}$ of $E_7$, and it can be seen that any other exceptional representations can be obtained by a repeated application of the decoupling procedure, see Table~\ref{decoupling}. 
Namely, given a geometry for the representation $R$ of $G$, one can give a vev $\vev{\phi}$ breaking $G$ to $G'$. At the same time, we can tune the mass $M$ of the hypermultiplet in $R$ of the order $M\sim \vev{\phi}$ so that the mass of the hypermultiplet in a subrepresentation $R'$ of $G'$ remains finite. 
This decoupling method produces  $X_{R'}$ for $G'$ given $X_R$ for $G$, which we illustrate in an example in Sec.~\ref{sec20}.
Now,  $X_{\mathbf{56}}$ of $E_7$ was determined in \cite{Hashiba:1999ss}.
Therefore, every other $X_R$ for exceptional representations follow. 

\begin{table}
\[
\begin{array}{ccccccccc}
&&E_7, \mathbf{56} \\
&&\downarrow & \searrow\\
&&E_6,\mathbf{27} & &\SO(12), \mathbf{32s} \\
&\swarrow& &\swarrow &\downarrow \\
\SU(6), \mathbf{15} &&  \SU(6), \mathbf{20} && \SO(10), \mathbf{16} \\
&\swarrow&&&\downarrow \\
\SU(5), \mathbf{10}&&&& \SO(8), \mathbf{8s} \\
&&&&\downarrow\\
&&&&\SU(4), \mathbf{4} 
\end{array}
\]
\caption{Relations among  `nice'  exceptional representations under decoupling \label{decoupling}}
\end{table}

Third, suppose we find such polynomials $X_{R_1}$  and $X_{R_2}$ for two representations $R_1$ and $R_2$ of $G$. Then the geometry for the combined representation can be easily found: \begin{equation}
\Lambda^{h^\vee-b_{R_1}-b_{R_2}} z X_{R_1} X_{R_2} =W_G \label{combined}
\end{equation} Indeed, defining $\tilde z= z \Lambda^{-b_{R_2}} X_{R_2}$, we get just $\Lambda^{h^\vee-b_{R_1}} \tilde z X_{R_1}=W_G$.  Therefore, assuming a three-cycle representing the hypermultiplets in $R_1$ is known in the latter geometry, the same three-cycle can be found in the geometry \eqref{combined}. Therefore the geometry for the reducible representations can be found by multiplying $X_R$ for the irreducible representations.

\subsection{General recipe}
Now we come to a general method to write down the Seiberg-Witten
geometry for $\cN=2$ theory with simply-laced gauge group $G$ with
matter content $\oplus_i R_i$, such that $b_{R_i}\le h^\vee$ to have
$X_R$ and $\sum_i b_{R_i} \le 2h^\vee$ to be asymptotically free or conformal. 
We split the irreducible components into two groups, $\oplus_a R_a$ and $\oplus_{\tilde a} \tilde R_{\tilde a}$ so that $\sum_a b_{R_a}\le h^\vee$ and $\sum_{\tilde a} b_{\tilde R_a}\le h^\vee$. Then we consider the geometry \begin{equation}
\Lambda^{h^\vee} z \prod_a \Lambda^{-b_{R_a}} X_{R_a} +
 \frac{\Lambda^{h^\vee}}z  \prod_{\tilde a}\Lambda^{-b_{\tilde R_{\tilde
 a}}} X_{\tilde R_{\tilde a}} = W_G(x_1,x_2,x_3;w_k). \label{swgeom}
\end{equation} When $\Lambda $ is very small, the geometry is of the form shown in Fig.~\ref{geometry} a), and thus we find three-cycles representing W-bosons and monopoles, together with three-cycles representing hypermultiplets in $\oplus_i R_i$ in the regions $|z|\sim 0$ and $|z|\sim \infty$. 
At least the one-loop running is reproduced by construction.  We can hope that the holomorphy guarantees that the geometry is correct even in the strongly-coupled regime.

We needed to divide $\oplus_i R_i$ into two subsets, but the resulting geometry in fact does not depend on the choice, because it can be easily rewritten to \begin{equation}
z + \frac{\Lambda^{2h^\vee}}z  \prod_{\tilde a}\Lambda^{-b_{\tilde R_{\tilde a}}} X_{\tilde R_{\tilde a}} = W_G.
\end{equation}

Admittedly, there are many gaps in the rough argument presented above. In the next section, we provide detailed checks of the construction in three examples.

We note that 
the Seiberg-Witten geometry (\ref{swgeom}) can be always rewritten in
the form 
\begin{equation}
W_G(x'_1,x'_2,x'_3;\tilde{w})=0, \label{tw}
\end{equation}
by a redefinition $x'_i=x'_i(x_k,w_k,m,\Lambda,z)$ and 
$\tilde{w}_k=\tilde{w}_k(w_k,m,\Lambda,z)$.\footnote{
Below, we will simply denote $\tilde{w}_k(w_k,m,\Lambda,z)$ as $w_k(z)$.}
Then, we can construct a corresponding Seiberg-Witten curve:
\begin{equation}
P_R(x,\tilde{w}_i)=0
\end{equation}
from the deformed Casimirs $\tilde{w}_i$ defined by (\ref{tw}) and
the characteristic polynomial of a representation $R$ of $G$,
$P_R(x,w_i)=\prod_\alpha (x-a \cdot \lambda_\alpha)$, where
$\lambda_\alpha$ is a weight of $R$.
We expect that
the low energy effective actions computed from this curve and the Seiberg-Witten geometry
are the same. 
The equivalence was explictly verified by \cite{Lerche:1996an} for pure $E_6$ theory, 
but it holds universally. 

In particular, the physics does not depend on the choice of the
representation $R$ as discussed in \cite{Hollowood:1997pp}, in which 
the cycles to be integrated on were explicitly specified.
Thus, we can use the curve instead of the geometry if we want to do so.
The Seiberg-Witten differential will be given by\footnote{
Precisely speaking, this is not completely fixed by the considerations so far
partially because there are ambiguities for the definition of
$\tilde{w}_i$. We will explain how to fix it in the Appendix~\ref{cg}.}
\begin{equation}
\lambda=x \frac{dz}{z}.
\end{equation}
These two descriptions are related in string theory by
the duality between the IIB string on non-compact Calabi-Yau (\ref{swgeom}) and 
the 6d (2,0) theory (or the M5-branes in the non-trivial background) on the curve.

\section{Case studies}\label{comparison}
\subsection{2-index antisymmetric tensor of $\SU(n)$}
\paragraph{Derivation} The curve of $\SU(n)$ with two hypermultiplets in the 2-index antisymmetric tensor representation with equal mass $m$ was determined in \cite{Argyres:2002xc}\footnote{Their coordinate system and ours are related as 
$z_\text{theirs}=u_\text{ours}$,
$y_\text{theirs}/\hat x_\text{theirs}=x_\text{ours}$, 
$\hat x_\text{theirs}=z_\text{ours}/\Lambda^2$, and $m_{A,\text{theirs}}=-2m_\text{ours}$. We set $\hat w_\text{theirs}=1$. 
Then our \eqref{ambient} is their (5.52), our \eqref{curve} is their (5.53), and our \eqref{RS} is their (5.40).
Their (5.53) has $x-2m_A$ instead of $x$ in \eqref{curve}, but that was a typo. The authors thank P.~Argyres for correspondences on this point.}: 
\begin{align}
\Lambda^2 (z+\frac 1z)  & = x^2 - u, \label{ambient}\\
0 &= R(u)+x S(u) \label{curve}
\end{align} where polynomials $R(u)$ and $S(u)$ are determined by the condition \begin{equation}
R(x^2)+x S(x^2)= \prod_i (x-a_i+\frac m2),\label{RS}
\end{equation}  and the SW differential is given by \begin{equation}
\lambda= x\frac{dz}z.
\end{equation} 
This curve is equivalent to the following non-compact Calabi-Yau, as we will show momentarily: \begin{equation}
\Lambda^2(z+\frac 1z) X_\text{asym}(x_1,x_2,x_3;w_i;m) = x_2 x_3 +P(x_1;w_i) \label{asymgeom}
\end{equation} where \begin{align}
P(x_1;w_i)&=\prod_i (x_1-a_i),\label{P}\\
X_\text{asym}(x_1,x_2,x_3;w_i;m) &= -x_3 + T(x_1,x_2;w_i;m).
\end{align}  Here, the polynomial $T(x_1,x_2;w_i;m)$ is defined via the relation \begin{equation}
x_2 T(x_1,x_2) = R((x_1-\frac m2)^2+x_2)+(x_1-\frac m2) S((x_1-\frac m2)^2 + x_2)-P(x_1). \label{Tdef}
\end{equation} Note that the right hand side has a factor of $x_2$ due to \eqref{RS} and \eqref{P}, guaranteeing that $T(x_1,x_2)$ is a polynomial.

Let us show the equivalence of the Calabi-Yau and the curve. Note that  $x_3$ only appears linearly in \eqref{asymgeom} and can be `integrated out'. The derivative with respect to $x_3$ of \eqref{asymgeom} sets \begin{equation}
-x_2=\Lambda^2(z+\frac1z). 
\end{equation} Plugging it back to \eqref{asymgeom} and using \eqref{Tdef}, we find \begin{equation}
R((x_1-\frac m2)^2+x_2)+(x_1-\frac m2) S((x_1-\frac{m}2)^2+x_2)=0.
\end{equation} Introducing $x=x_1-\frac m2$ and $u=(x_1-\frac m2)^2+x_2$, we find the curve \eqref{ambient} and \eqref{curve}.

\paragraph{Confirming the discriminant}
From the discussion above, we conclude that the hypermultiplet geometry representing a 2-index antisymmetric representation of $\SU(n)$ is given by \begin{equation}
\Lambda^2 z X_\text{asym}(x_1,x_2,x_3;w_i;m)=x_2 x_3 +P(x_1;w_i). \label{asymhyper}
\end{equation} Let us check that this geometry becomes singular when a component of the hypermultiplet becomes massless. The geometry \eqref{asymhyper} becomes singular when its derivatives with respect to $x_1,x_2,x_3$ all becomes zero. One easily sees that this is equivalent to the fact that the equations $R(u)=0$ and $S(u)=0$ defined in \eqref{RS} have a common zero $u=u_0$, i.e.~$R(u_0)=S(u_0)=0$. This happens when their resultant is zero; and in general the resultant is given by \begin{equation}
\text{Resultant of $R(u)$ and $S(u)$}=\text{const}.\prod_{i<j} (a_i+a_j - m). \label{resultant}
\end{equation} Therefore, the geometry becomes singular when a component of the antisymmetric tensor becomes massless. 
To confirm the resultant, assume $a_i+a_j=m$ for a pair $i<j$. Let $x_0=a_i-m/2=-a_j+m/2$. Using \eqref{RS}, we find 
$R(x_0^2)+x_0 S(x_0^2)=R(x_0^2)-x_0 S(x_0^2)=0$. $x_0$ is generically nonzero, so we conclude $R(x_0^2)=S(x_0^2)=0$. 
This means that the resultant has a factor of $a_i+a_j-m$ for each pair $i<j$. One can independently calculate the degree of the resultant because we know the degrees of $R(u)$ and $S(u)$, and we conclude that there is no other factor.

\subsection{3-index antisymmetric tensor $\mathbf{20}$ of $\SU(6)$}\label{sec20}
\paragraph{Decoupling from $\mathbf{32}$ of $\SO(12)$ }
One representation for which $X_R$ has not been written down is the three-index antisymmetric tensor $\mathbf{20}$ of $\SU(6)$.\footnote{There was a paper \cite{Rhedin:2000bc} in which it was attempted to determine the Seiberg-Witten curve for gauge theories with three-index antisymmetric tensors.} This can be derived from the known case, $\mathbf{32}$ of $\SO(12)$ via decoupling.  We start from $X_{\mathbf{32}}$, obtained in \cite{Hashiba:1999ss}:
\begin{multline}
X_{\mathbf{32}}(x_i;W_i;m)=\frac{1}{256}(8 i x_2 + 8 x_1^2 + 4 W_2 x_1 + 4 W_4- W_2^2 )^2 \\
+m^2(i\tilde W_6 + W_6 - \frac14 W_2 W_4 + \frac 1{16} W_2^3+2W_2 x_1^2 -\frac 18 W_2^2 x_1 + \frac32 W_4 x_1 + 3 x_1^3 + 3i x_1 x_2 + \frac i2 W_2 x_2) \\
+4 i m^3 x_3 + m^4( 3i x_2 + \frac12 W_2 x_1 + \frac38 W_2^2-\frac12 W_4) + (2 x_1 + W_2)m^6+ m^8.
\end{multline}
With this, we consider the geometry
\begin{equation}
\Lambda^{10-8} 
z X_{\mathbf{32}}=x_1^{5} + x_1 x_2^2 - x_3^2  + W_2 x_1^{4}+ W_4 x_1^{3} + W_6 x_1^2 + W_8 x_1 + W_{10}  + \tilde W_6.
\end{equation} 

Now we give a vev to $W_i$ so that $\SO(12)$ is broken to $\SU(6) \times
U(1)$. The spin representation decomposes as \begin{equation}
\mathbf{32}\to \mathbf{20}_0 \oplus \mathbf{6}_1\oplus \bar{\mathbf{6}}_{-1}.
					 \end{equation} 
The vev gives the mass to the hypermultiplets proportional to 
the $U(1)$ charges through the $ \sqrt{2} \tilde{Q} \Phi Q$ term.
Therefore, if we take the vev infinite with 
a shift of the mass $m$ to cancel this additional contribution,
we will have $SU(6)$ gauge theory with $\mathbf{20}$ or $\mathbf{6}$,
depending on the shift. Here, we should rescale the dynamical scale 
using the scale matching condition as usual and the $U(1)$ sector decouples by the limit.
In order to obtain $\mathbf{20}$ we do not need to shift the mass and 
just give the vev and take the limit.
The geometry in the limit will be obtained 
with a non-trivial redefinition of the coordinates 
$x'_i=x'_i(x_1,x_2,x_3,m,W_k)$
such that the geometry is written as
\begin{equation}
W_{SO(12)}(x_1,x_2,x_3,W_k)=M^4 W_{SU(6)}(x'_1,x'_2,x'_3,w_k)+{\cal O}(M^3),
\end{equation}  
where $M$ is the scale of the vev.
This procedure was explained in detail in \cite{Terashima:1998fx}.
Explicitly doing this with the help of a computer, 
we finally obtain the geometry for $\mathbf{20}$ of $\SU(6)$: \begin{equation}
\Lambda^{6-6} zX_{\mathbf{20}} = x_2 x_3 + x_1^6 + w_2 x_1^4 + w_3 x_1^3+ w_4 x_1^2 + w_5 x_1 + w_6 \label{bosh}
\end{equation} where 
\begin{multline}
X_{\mathbf{20}}(x_i;w_i;m)=-(w_2 x_1+w_3 + 2x_1^3+2 i x_2)^2 \\
+ m^2 (-12 x_1^4-12 i x_1 x_2 - 6 w_2 x_1^2- 6 w_3 x_1 - 4 w_4 + w_2^2) \\
 + 8 m^3 x_3 + m^4 (3 x_1^2+2 w_2) + m^6. 
\end{multline} It is relieving to find that it becomes a square when $m$ is set to zero, because $\mathbf{20}$ is pseudo-real and one can consider a half-hypermultiplet in $\mathbf{20}$: 
\begin{equation}
X_{\frac12\mathbf{20}}(x_i;w_i)=i(w_2 x_1+w_3 + 2x_1^3+2 i x_2).
\end{equation}

\paragraph{Comparison with 6d $\cN=(2,0)$ theory: introductory remarks} 
Let us study this geometry from the point of view of the 6d $\cN=(2,0)$ theory of type $A_5$. 
It is done by rewriting the geometry in the form \begin{equation}
 x_2 x_3 + x_1^6 + w_2(z) x_1^4 + w_3(z) x_1^3+ w_4(z) x_1^2 + w_5(z) x_1 + w_6(z) =0 \label{aho}
\end{equation} and considering the multi-differentials \begin{equation}
\varphi_k(z)= w_k(z) \frac{dz^k}{z^k}
\end{equation} as the worldvolume fields of the 6d theory. 
A further compactification on $S^1$ makes the system into 5d maximally supersymmetric Yang-Mills on a sphere parameterized by $z$. Then there is a complex adjoint scalar field $\Phi(z)$ which is a differential such that \begin{equation}
\lambda^6 + \varphi_2(z) \lambda^4 + \cdots + \varphi_6(z) = \det(\lambda- \Phi(z))
\end{equation} where $\lambda$ is the Seiberg-Witten differential. 

The main point of the approach using the adjoint field $\Phi(z)$ \cite{Gaiotto:2009we,Gaiotto:2009hg} is to identify its singularity. This method is most developed when the total $b$ equals $h^\vee$. In this case, it is expected that there are three singularities, say at $z=0,1,\infty$. Let $t$ be a local coordinate at a singularity so that the singularity is at $t=0$. Then $\Phi$ has the form  \begin{equation}
\Phi \sim  \Phi_{-1} \frac{dt}t + \text{regular}
\end{equation} with a residue $\Phi_{-1}$. 
The residue $\Phi_{-1}$ is diagonal when the Seiberg-Witten differential $\lambda$ has a single pole there, representing the hypermultiplet mass term. Even when the hypermultiplet mass term is zero, $\Phi_{-1}$ can be a nonzero nilpotent matrix. The Jordan decomposition of this nilpotent matrix captures the important data of a singularity. There is now a method of reproducing the number of hypermultiplets and its flavor symmetry given the type of three singularities \cite{Gaiotto:2009we,Chacaltana:2010ks}.

\paragraph{Half $\mathbf{20}$ and three fundamentals}
To apply this technology, first take a half-hypermultiplet of $\mathbf{20}$ and add three copies of fundamentals so that the total $b$ is $6$: \begin{multline}
z(x_1-m_1)(x_1-m_2)(x_1-m_3)i(w_2 x_1+w_3 + 2x_1^3+2 i x_2) = \\
x_2 x_3 + x_1^6 + w_2 x_1^4 + w_3 x_1^3+ w_4 x_1^2 + w_5 x_1 + w_6,
\end{multline} which can be readily made into the form \eqref{aho}. 
One finds the following:
\begin{itemize}
\item  at $z=0$, $\Phi_{-1}$ is a generic diagonal matrix. This puncture is called the full puncture.
\item  at $z=1$, $\Phi_{-1}\propto \diag(1,1,1,1,-2,-2)$ when $m_1+m_2+m_3\ne 0$.
When $m_1+m_2+m_3=0$, the order $p_k$ of the pole of $\varphi_k$ is given by $(p_2,p_3,p_4,p_5,p_6)=(1,1,2,2,2)$, which corresponds to a nilpotent $\Phi_{-1}=N_2\oplus N_2\oplus N_1\oplus N_1$ where $N_c$ is a $c\times c$ Jordan block.  This corresponds to the puncture labeled by a partition $[4,2]$.
\item  at $z=\infty$, $\Phi_{-1}\propto \diag(a,a,b,b,c,c)$ where $(a,b,c)$ is the traceless part of $(m_1,m_2,m_3)$. When it is zero,  the orders of the poles of $\varphi_k$ are given by $(p_2,p_3,p_4,p_5,p_6)=(1,2,2,3,4)$, which corresponds to a nilpotent $\Phi_{-1}=N_3\oplus N_3$.   This corresponds to the puncture labeled by a partition $[2,2,2]$.
\end{itemize}
Using the method explained in \cite{Chacaltana:2010ks}, it is straightforward to find that this three-punctured sphere has the requisite properties to be identified with $\frac12\mathbf{20}$ plus three copies of $\mathbf{6}$.
For example, let us check that the $\cN=2$ system defined by this three-punctured sphere does not have any Coulomb branch, as is required for any theory consisting only of hypermultiplets.
In general, the dimension of the Coulomb branch of a sphere with many punctures is obtained by the formula \cite{Gaiotto:2009we,Chacaltana:2010ks,Benini:2010uu} \begin{equation}
\dim(\text{Coulomb branch})=\sum_{i} \dim_\bC O_G(\Phi_{-1,i}) - 2 \dim G\label{CoulombDim}
\end{equation} where $\Phi_{-1,i}$ is the residue at the $i$-th singularity and $O_G(\Phi)$ is the space of elements conjugate to $\Phi$ under $G_\bC$.
In our current situation, we have \begin{equation}
\dim O(\text{generic})=30,\quad
\dim O(N_2\oplus N_2 \oplus N_1\oplus N_1) = 16,\quad
\dim O(N_3\oplus N_3) = 24.\quad
\end{equation} Then the dimension of the Coulomb branch is \begin{equation}
\dim(\text{Coulomb branch})=30+16+24-2\cdot 35=0,
\end{equation} which is zero as it should be.

\paragraph{Full $\mathbf{20}$}
Next, consider the full half-hypermultiplet $\mathbf{20}$, which in itself satisfy $b_{\mathbf{20}}=6$. We rewrite the geometry \eqref{bosh} into \eqref{aho}. We find three singularities of $\varphi_k$: \begin{itemize}
\item at $z=0$, one has again the full puncture, with the flavor symmetry $\SU(6)$.
\item at $z=1$, one finds that $\varphi_{3}$ and $\varphi_{5}$ have the branch cut of the form $\sim (z-1)^{1/2}$.   We find that the orders of poles are given by $(p_2,p_3,p_4,p_5,p_6)=(1,\frac32,3,\frac52,3)$, but by redefining the Casimirs as  \begin{equation}
\varphi'_2=\varphi_2,\ 
\varphi'_3=\varphi_3,\ 
\varphi'_4=\varphi_4-\frac14\varphi_2{}^2,\ 
\varphi'_5=\varphi_5-\frac12\varphi_2\varphi_3,\ 
\varphi'_6=\varphi_6-\frac14\varphi_3{}^2,\label{better}
\end{equation} we find the poles are given by $(p_2',p_3',p_4',p_5',p_6')=(1,\frac32,1,\frac32,2)$.
\item at $z=\infty$, one similarly finds that $\varphi_{3}$ and $\varphi_{5}$ have branch cuts.
The orders of poles in the massless case are given by $(p_2,p_3,p_4,p_5,p_6)=(1,\frac32,3,\frac72,4)$, which can not be further reduced by redefinitions of Casimirs.  When the hypermultiplet is massive, we have \begin{align}
\varphi_2(z)&\sim -3m^2 \frac{(dz)^2}{z^2},&
\varphi_3(z)&\sim  \frac{(dz)^3}{z^{3/2}},&
\varphi_4(z)&\sim  +3m^4\frac{(dz)^4}{z^{4}},\\
\varphi_5(z)&\sim  \frac{(dz)^5}{z^{7/2}}, &
\varphi_6(z)&\sim  -m^6\frac{(dz)^6}{z^{6}}.
\end{align}
\end{itemize}

The presence of branch cuts means that the 6d construction involves the $\bZ_2$ outer automorphism, as in \cite{Tachikawa:2010vg}.
This means that the Hitchin field is transposed, $\Phi(z)\to \sigma(\Phi(z))$, when we go around the puncture.
Here $\sigma(X)$  can be \begin{equation}
\sigma(X) =-X^t.
\end{equation} But this choice of $\sigma$ does not leave any diagonal matrix invariant, and thus inconvenient.
Instead let us use $\sigma$ defined by \begin{equation}
\sigma(E_{i,j})= - (-1)^{i+j} E_{7-j,7-i}
\end{equation} where $E_{i,j}$ is the matrix with 1 at $(i,j)$-th entry and zero otherwise.
Then the diagonal matrix of the form \begin{equation}
\diag(a,b,c,-c,-b,-a)
\end{equation} is preserved; we can check that the twist $\sigma$ preserves $\USp(6)$ subgroup of $\SU(6)$.
 So, when we have branch cuts, we expect $\Phi$ to behave as \begin{equation}
\Phi(t)\sim \Phi_{-1} \frac{dt}t + \Phi_{-1/2} \frac{dt}{t^{1/2}} +
 \Phi_{0} \frac{dt}{t^0} 
+ \text{lower order terms}
\end{equation}
so that $\sigma(\Phi_{-1})=\Phi_{-1}$, $\sigma(\Phi_{-1/2})=-\Phi_{-1/2}$
and $\sigma(\Phi_0)=\Phi_0$. 
The leading term $\Phi_{-1}$ determines the property of the puncture, and $\Phi_{-1/2}$ and $\Phi_0$ will be generic. 
We find that \begin{itemize}
\item at $z=1$, the choice $\Phi_{-1}=0$ reproduces the behavior of $\varphi_k$.
\item at $z=\infty$, we can reproduce the behavior of $\varphi_k$ by choosing $\Phi_{-1}=N_2\oplus N_2\oplus N_2$ when $m=0$, 
and $\Phi_{-1}\propto \diag(1,1,1,-1,-1,-1)$ when $m\ne 0$. These massive and massless $\Phi_{-1}$ are the one associated to the partition $[3,3]$ if there is no branch cut.
\item at $z=0$, we do not have any branch cut, and $\Phi_{-1}$ is just a generic diagonal matrix. 
\end{itemize}
Thus, at least we have checked that our polynomial $X_R$ can come from a singularity of the Hitchin field $\Phi$.  
Let us check that the $\cN=2$ theory defined by these punctures does not have the Coulomb branch, as it should be for a theory of free hypermultiplets. 
In our situation, the formula \eqref{CoulombDim} cannot be directly used, because some of the punctures have branch cuts. In this case we need to use the formula \cite{Benini:2010uu} \begin{equation}
\dim(\text{Coulomb branch})=\sum_{i} \dim_\bC O_{G_i}(\Phi_{-1,i}) - 2 \dim G_\text{common}
\end{equation} where $G_i$ is the gauge group preserved by the twist at the $i$-th puncture, and $G_\text{common}$ is the common subgroup of all $G_i$. 
To apply this formula to our current case, we use \begin{equation}
O_{\SU(6)}(\text{generic})=30,\quad 
O_{\USp(6)}(0)=0,\quad 
O_{\USp(6)}(N_2\oplus N_2\oplus N_2)=12
\end{equation} and $G_\text{common}=\USp(6).$ Then \begin{equation}
\dim(\text{Coulomb branch})=30+0+12 - 2\cdot 21 =0,
\end{equation} as it should be.

It is not yet developed how the properties of the hypermultiplet can be recovered the choice of the singularities of the Hitchin field in the presence of the branch cuts of 6d $A_{n-1}$ theory, so we cannot perform further checks in this case.

\paragraph{Six fundamentals} As a comparison, consider six fundamentals, which also have total $b=6$. The geometry is \begin{equation}
z \prod_{i=1}^6 (x-m_i) = x_2^2 + x_3^2 + x_1^6 + w_2 x_1^4 + w_3 x_1^3+ w_4 x_1^2 + w_5 x_1 + w_6,
\end{equation} which can be made into the form \eqref{aho} very easily. One finds \cite{Gaiotto:2009we} \begin{itemize}
\item At $z=0$ and $z=\infty$, one finds the full puncture, each carrying $\SU(6)$ flavor symmetry.
\item At $z=1$, the poles behave as $(p_2,p_3,p_4,p_5,p_6)=(1,1,1,1,1)$. The residue of the Hitchin field is given by $\Phi_{-1}\propto M(1,1,1,1,1,-5)$ where $M=\sum m_i$ in general; it becomes $\Phi_1 = N_2\oplus N_1\oplus N_1\oplus N_1 \oplus N_1$ when $M=0$. This singularity is usually called the simple puncture.
\end{itemize}

\paragraph{Summary} Thus, we saw the three cases, 
i) half $\mathbf{20}$ and three fundamentals,
ii) full $\mathbf{20}$,
iii) six fundamentals,
all have realizations in terms of 6d $\cN=(2,0)$ theory of type $A_5$ on three-punctured spheres. 
We have found that the choice of punctures in each case was completely different from each other; the second case involved even a $\bZ_2$ outer automorphism. 
These diverse examples, however, all came from the same ingredient $X_{\mathbf{20}}$ and $X_{\mathbf{6}}$ within the approach of this paper.

\begin{figure}
\[
\includegraphics[width=.8\textwidth]{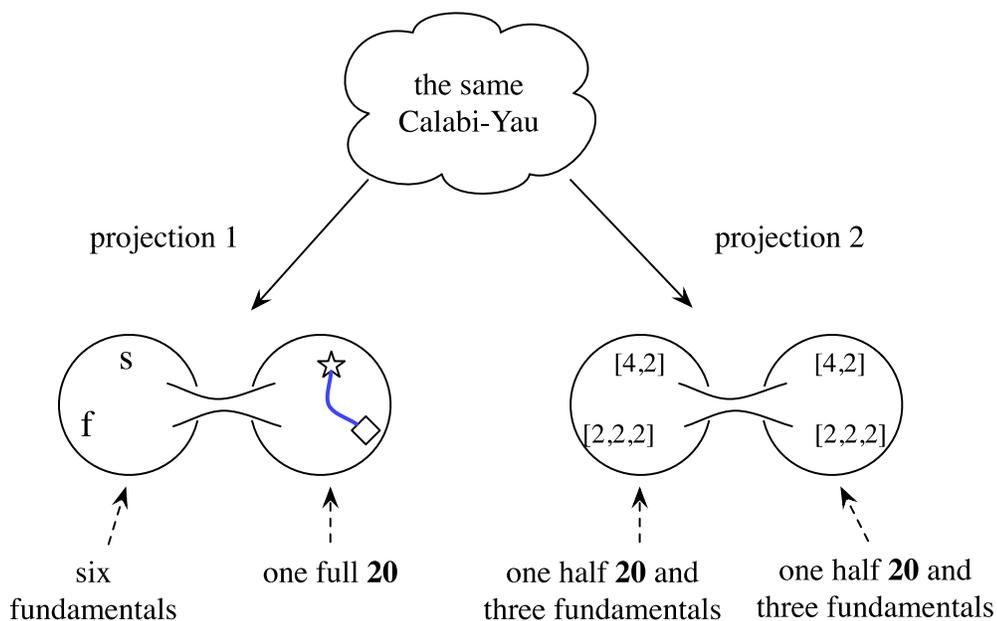}
\]
\caption{The same non-compact Calabi-Yau geometry \eqref{same} leads to two completely different sets of codimension-two ``punctures'' as 6d $\cN=(2,0)$ theory on a sphere. On the left, we chose the projection to the sphere parameterized by $z'$, \eqref{z'}. The pair of a simple puncture and a full puncture gives us six fundamentals, and the pair of two punctures with $\bZ_2$ branch cuts produces one full hypermultiplet in $\mathbf{20}$.
On the right, the projection was to the sphere parameterized by $z''$, \eqref{z''}. Two pairs of a puncture of type $[4,2]$ and another of type $[2,2,2]$ each give one half-hypermultiplet in $\mathbf{20}$ and three fundamentals.
\label{projections}}
\end{figure}

As another intriguing consequence of this analysis, let us consider superconformal $\SU(6)$ gauge theory with one $\mathbf{20}$ and six fundamentals. The Seiberg geometry is given by  \begin{equation}
z+\frac{q}z X_{\mathbf{20}} X_{\mathbf{6}}{}^6 = W_{A_5}. \label{same}
\end{equation} One can redefine $z$ to rewrite the geometry in the form \begin{equation}
z' X_{\mathbf{6}}{}^6+\frac{q}{z'} X_{\mathbf{20}}  = W_{A_5}.\label{z'}
\end{equation} As a six-dimensional theory living on the sphere parameterized by $z'$, we have a simple puncture and a full puncture on the left giving six fundamentals, and two punctures with branch cuts on the right giving a full $\mathbf{20}$. 
One can also redefine $z$ in such a way to have the geometry of the form \begin{equation}
z'' X_{\mathbf{6}}{}^3X_{\frac12\mathbf{20}} +\frac{q}{z''} X_{\mathbf{6}}^3X_{\mathbf{20}}  = W_{A_5}.\label{z''}
\end{equation}
Then, the 6d theory living on the sphere parameterized on $z''$ has, on both sides, one puncture of type $[4,2]$ and another of $[2,2,2]$.
This illustrates the fact that the same Seiberg-Witten geometry \eqref{same} can lead to two completely different combinations of codimension-two defects on the 6d theory, when the choice of the base was changed from $z'$ to $z''$, see Fig.~\ref{projections}.

\subsection{$\mathbf{56}$ of $E_7$}\label{sec56}
Let us next consider $\mathbf{56}$ of $E_7$, whose polynomial is given by \begin{multline}
X_{\mathbf{56}}(x_i;w_i;m)= x_2^2 + m^2(-6 x_1 x_2 - 4 w_{10}) + 8im^3 x_3 \\
+m^4(-3x_1^2-6w_2 x_2-4w_8) + m^6(2w_2x_1 -10 x_2 -4 w_6) \\
+m^8(6x_1+w_2^2) + 2 w_2 m^{10}+m^{12} .
\end{multline} For a half-hypermultiplet, the polynomial is very simple, $X_{\frac12\mathbf{56}}=x_2$.

\paragraph{Behavior of Casimirs}
The total $b$ becomes $h^\vee=18$ if we use three half-hypermultiplets. The geometry is then \begin{multline}
z x_2 X_{\mathbf{56}}(x_i;w_i;m) = x_1^3{x_2} +x_2^3+x_3^2 \\
 + w_2 {x_2}^2 {x_1} + w_6 {x_2}^2  + w_8 {x_2} {x_1} + w_{10} {x_1}^2 + w_{12} {x_2} + w_{14} {x_1}+ w_{18}. 
\end{multline} Let us study this geometry from the viewpoint of $\cN=(2,0)$ theory of type $E_7$. 
We first rewrite it in the form \begin{multline}
0=x_1^3{x_2} +x_2^3+x_3^2 
 + w_2(z) {x_2}^2 {x_1} + w_6(z) {x_2}^2 \\ + w_8(z) {x_2} {x_1} + w_{10}(z) {x_1}^2 + w_{12}(z) {x_2} + w_{14}(z) {x_1}+ w_{18}(z), 
\end{multline} and consider $\varphi_k(z)=w_k(z) (dz/z)^k$ as the worldvolume fields.
We find the following singularities: \begin{itemize}
\item at $z=0$, we have a singularity where $\varphi_k(z) \sim w_k (dz/z)^k$. 
\item at $z=1$, the order of the poles of $\varphi_k(z)$ is given by \begin{equation}
(p_2,p_6,p_8,p_{10},p_{12},p_{14},p_{18})=(1,2,2,2,3,3,4). \label{z1pole}
\end{equation} This cannot be further reduced by redefinition of the Casimirs. This does not depend on $m$, either.
\item at $z=\infty$ with $m\ne 0$, the leading order of $\varphi_k(z)$ behaves as \begin{align}
w_2(t) &\sim -6m^2,  &
w_6(t) &\sim +10m^6,&
w_8(t) &\sim -3m^8,&
w_{10}(t) &\sim 0 ,\nonumber \\
w_{12}(t)&\sim -2m^{12},&
w_{14}(t)&\sim 0, &
w_{18}(t)&\sim 0 \label{zinfmassive}
\end{align}  where $t=1/z$. $\varphi_k(z)$ is obtained by multiplying $w_k(t)$ by $(dt/t)^k$.
When $m=0$,  the order of the poles is given by  \begin{equation}
(p_2,p_6,p_8,p_{10},p_{12},p_{14},p_{18})=(1,4,6,8,9,11,14). \label{zinfpole}
\end{equation}
\end{itemize}

\paragraph{Behavior of the Hitchin field}
Any one of these behaviors should come from a singularity of the Hitchin field, $\Phi(t)\sim \Phi_{-1} dt/t + \text{regular}$. Indeed, by trial and error, one finds that the following choices do the job: \begin{itemize}
\item at $z=0$, we just have to take $\Phi_{-1}$ to be a general Cartan element $h$: \begin{equation}
\Phi(z)= h \frac{dz}z + \text{regular}.
\end{equation}
 This is the full puncture.
\item at $z=1$, we can take $\Phi_{-1}$ to be a raising operator $N_\alpha$ corresponding to any root $\alpha$: \begin{equation}
\Phi(t)= N_\alpha \frac{dt}t + \text{regular}
\end{equation}  where $t=z-1$. The pole behavior \eqref{z1pole} was
      checked by choosing a random regular element, taking
      $\tr\Phi(t)^k$ for $k=2,\ldots, 18$, and finally extracting the
      Casimirs using the formulas in
      \cite{KatzMorrison,Shioda}\footnote{The structure constants of any
      simple Lie algebra are available in electronically readable form 
in the  \href{http://www.gap-system.org/Datalib/lie.html}{Lie algebra package} 
of \href{http://www.gap-system.org/}{GAP}. 
The authors thank David Vegh for the help.}. 
This orbit $N_\alpha$ is known to be rigid, i.e.~cannot be deformed to include mass terms, see e.g.~Appendix A of \cite{Gukov:2008sn}.
\item at $z=\infty$, we can take $\Phi_{-1}$ to be the fundamental weight vector $v_\text{center}$ corresponding to the central node of $E_7$ Dynkin diagram where three legs are joined: \begin{equation}
\Phi(t)= mv_\text{center} \frac{dt}t + \text{regular}.
\end{equation} This can be readily translated to the behavior of $w_k(t)$ using the formula given in \cite{KatzMorrison}, reproducing \eqref{zinfmassive}. When the mass parameter $m$ is turned off, the adjoint element conjugate to $mv_\text{center}$ is known (see e.g.~\cite{Moreau}) to degenerate to a nilpotent element $N_{A_4+A_2}$ of the Bala-Carter label $A_4+A_2$: \begin{equation}
\Phi(t)= N_{A_4+A_2} \frac{dt}t + \text{regular}.
\end{equation} The Bala-Carter label means the following: one takes a subgroup of the corresponding type, \begin{equation}
\SU(5)\times \SU(3) \subset E_7,
\end{equation} which is apparent from the Dynkin diagram. Then, we take the biggest Jordan blocks in the subgroup: \begin{equation}
N_5\oplus N_3 \in \mathrm{su}(5)\oplus \mathrm{su}(3) \subset E_7.
\end{equation}  Then $N_{A_4+A_2}=N_5\oplus N_3$ is the same element regarded as an element of $E_7$. Again, the behavior of the poles can be computed and it reproduces \eqref{zinfpole}.
\end{itemize}

Thus we identified the type of the three punctures on the sphere, on which 6d $\cN=(2,0)$ theory is compactified. Let us perform one final check by confirming that the Coulomb branch of  this theory is free is zero dimensional. 
Again by referring to e.g.~\cite{Moreau}, we find \begin{equation}
\dim_\bC O(h)=126,\quad
\dim_\bC O(N_\alpha)=34,\quad
\dim_\bC O(N_{A_4+A_2})=106.
\end{equation} Recalling $\dim E_7=133$ and using \eqref{CoulombDim}, we find \begin{equation}
\dim(\text{Coulomb branch})=126+34+106-2\cdot 133=0,
\end{equation} as it should be.

\section{Conclusions}\label{conclusions}
In this paper we argued that the Seiberg-Witten geometry of any single simply-laced gauge group with  matter content such that any irreducible representation $R$ in it satisfies $b_R\le h^\vee$ has a universal form given in \eqref{SWgeometry}. The basic point was that Type IIB string on the non-compact Calabi-Yau given by  \begin{equation}
z X_R(x_1,x_2,x_3;w_i;m) = W_G(x_1,x_2,x_3;w_i) \label{BAR}
\end{equation} represents the hypermultiplets in the representation $R$ of the group $G$.
We saw how the general geometry \eqref{SWgeometry} can be built from the basic ingredients of the form \eqref{BAR}.

We then analyzed these Seiberg-Witten solutions from the point of view of the 6d $\cN=(2,0)$ theory in  detail for three cases: the first was the 2-index antisymmetric tensor of $\SU(n)$, the second was  $\mathbf{20}$ of $\SU(6)$, and the third was  $\mathbf{56}$ of $E_7$. 
We studied the types of the punctures of 6d $\cN=(2,0)$ theory, starting from the equations of the non-compact Calabi-Yau, and identified  the residues of the adjoint one-form $\Phi$ on the sphere. 
We found that the Calabi-Yau geometries previously found via various methods can all consistently be interpreted as arising from the 6d $\cN=(2,0)$ theory on a sphere. 

There are a few obvious directions of further research. One is to extend our analysis to non-simply-laced gauge groups. This will involve $\bZ_2$ or $\bZ_3$ outer-automorphism, as was first seen in \cite{Martinec:1995by}. Another is to find the Seiberg-Witten solution to the `reasonable' cases, i.e.~when the matter content involves a representation $R$ in the range $h^\vee<b_R\le 2h^\vee$, see Table~\ref{repdata}.
The curves for the theories with massive adjoints or massive 2-index symmetric tensor of $\SU(n)$ are known \cite{Landsteiner:1997ei,D'Hoker:1998yi}.
So, the problematic ones are the 3-index antisymmetric tensor of $\SU(7)$ and $\SU(8)$, and the spinor of $\SO(14)$.  If the matter content is just one copy of one of these representations, then they can be obtained by starting from the $\cN=2^*$ theory with $E_7$ or $E_8$ gauge group and giving an appropriate vev to the adjoint scalar. Therefore, the curves for these cases are implicitly known.  The main problem would be to add additional matter fields in the  fundamental representation to these theories.

Finally, we need to note that although we have determined $X_R$ for all nice representations through various means, we do not yet have a direct understanding of the relation between the representation $R$ and the polynomial $X_R$.  
It would be desirable to have a more uniform, logical way which allows us to write down the polynomial $X_R$ given the weights of $R$.

\section*{Acknowledgements}
YT would like to thank the hospitality of the Yukawa Institute for the two weeks immediately after the earthquake;  this collaboration was initiated during that stay.
ST would like to thank Sung-Kil Yang from whom ST
learned many things through discussions 
and will not forget his warm encouragements.
S.T. would like to thank K. Hosomichi, K. Sakai and M. Taki for helpful discussions.
The  work of YT is supported in part by NSF grant PHY-0969448  and by the Marvin L. Goldberger membership through the Institute for Advanced Study, while he was there from April to July 2011.
The  work of YT is also supported in part by World Premier International Research Center Initiative (WPI Initiative),  he Japan Ministry of Education, Culture, Sports, Science and Technology (MEXT), Japan through the Institute for the Physics and Mathematics of the Universe, the University of Tokyo.
The work of ST is partly supported by MEXT, and by the Grant-in-Aid for the Global COE program ``The Next Generation of Physics, Spun from Universality and Emergence'' from the MEXT.

\appendix

\section{List of hypermultiplet geometries}\label{XR}

\subsection{$X_R$ for $\SU(n)=A_{n-1}$} 

The ALE space is given either by \begin{equation}
W_{A_{n-1}}=x_1^n+x_2^2+x_3^2  +  w_2 x_1^{n-2}+w_3 x_1^{n-3} + \cdots +w_n
\end{equation} or
\begin{equation}
W_{A_{n-1}'} = x_1^n+x_2 x_3  + w_2 x_1^{n-2}+w_3 x_1^{n-3} + \cdots +w_n.
\end{equation} 
The relation between $w_k$ and the Cartan element $\phi=\diag(a_1,\ldots,a_n)$ is standard: \begin{equation}
x^n + w_2 x^{n-2} + \cdots+ w_n=\prod_{i=1}^n (x-a_i).
\end{equation}

For the fundamental, we have \begin{equation}
X_\text{fund}(x_i ;w_i; m)=x_1-m.\label{fund}
\end{equation}  This formula goes back to \cite{Hanany:1995na,Argyres:1995wt}.

For the 2-index antisymmetric representation, we use $W_{A_{n-1}'}$ with \begin{equation}
X_\text{2-index antisym.}(x_i;w_i;m)=-x_3 +T(x_1,x_2;w_i;m)\label{asym}
\end{equation} where $T(x_1,x_2;w_i;m)$ is defined as follows. Let $P(x;w_i)=\prod_{i=1}^n (x-a_i)$, and define
two polynomials $R(u;w_i;m)$ and $S(u;w_i;m)$  via \begin{equation}
P(x+\frac m2;w_i)=R(x^2;w_i;m)+x S(x;w_i;m). 
\end{equation}  Then $T(x_1,x_2;w_i;m)$  is given by \begin{equation}
P(x_1;w_i)+x_2 T(x_1,x_2;w_i;m)= R((x_1-\frac m2)^2+x_2;w_i;m)+(x_1-\frac m2)^2 S((x_1-\frac m2)^2+x_2;w_i;m) 
\end{equation} These polynomials can be straightforwardly extracted from the curve given in \cite{Argyres:2002xc}. The curve was interpreted from the point of view of 6d $\cN=(2,0)$ theory in \cite{Nanopoulos:2009xe} and the punctures were identified there.
We can check that these polynomials for $n <7$ indeed coincide with the 
those obtained in \cite{Terashima:1998fx}.

For the 3-index antisymmetric representation $\mathbf{20}$ of $\SU(6)$, we use $W_{A_5}$ and \begin{multline}
X_{\mathbf{20}}(x_i;w_i;m)=-(w_2 x_1+w_3 + 2x_1^3+2 i x_2)^2 \\
+ m^2 (-12 x_1^4-12 i x_1 x_2 - 6 w_2 x_1^2- 6 w_3 x_1 - 4 w_4 + w_2^2) \\
 + 8 m^3 x_3 + m^4 (3 x_1^2+2 w_2) + m^6. \label{20}
\end{multline} For a half-hypermultiplet in this representation, we just set $m=0$ and take the square root: \begin{equation}
X_{\frac12\cdot\mathbf{20}}(x_i;w_i)=i (w_2 x_1+w_3 + 2x_1^3+2 i x_2).
\end{equation} The detailed analysis of this polynomial was given in Sec.~\ref{sec20}.

\subsection{$X_R$ for  $\SO(2n)=D_n$}

The ALE space is given by 
\begin{equation}
W_{D_{n}}  = x_1^{n-1} + x_1 x_2^2 - x_3^2  + w_2 x_1^{n-2}+ w_4 x_1^{n-3} + \cdots+ w_{2n-2} + \tilde w_n x_2.
\end{equation}
The relation between $w_k$, $\tilde w_n$ and the Casimirs $\phi=\diag(\pm a_i)$ is standard: \begin{equation}
x^n + w_2 x^{n-1} + w_4 x^{n-2} + \cdots + w_{2n-2} x + w_{2n} = \prod_{i=1}^n (x-a_i^2) 
\end{equation} and \begin{equation}
\tilde w_n = 2 i^{n+1} a_1 a_2\cdots a_n.
\end{equation}

For the vector representation, we have \begin{equation}
X_\text{vector}(x_i ;w_i; m)=x_1-m^2.\label{vector}
\end{equation}  This goes back to \cite{Argyres:1995fw,Hanany:1995fu}.

For the $\SO(8)$ spinor $\mathbf{8s}$, we have \begin{equation}
X_{\mathbf{8s}}(x_i;w_i;m)=\frac12 x_1 +\frac i2  x_2  +\frac14 w_2+m^2. \label{8s}
\end{equation} The polynomial  for the conjugate spinor $\mathbf{8c}$ can be obtained by applying the outer automorphism $x_2\to -x_2$: \begin{equation}
X_{\mathbf{8c}}(x_i;w_i;m)=\frac12 x_1 -\frac i2  x_2  +\frac14 w_2+m^2. \label{8c}
\end{equation} This was determined in \cite{Terashima:1998fx}.
The three-punctured spheres on which 6d $\cN=(2,0)$ theory of type $D_4$ is to be compactified to produce these matter contents were analyzed in \cite{Tachikawa:2010vg,Chacaltana:2011ze}; it would be interesting to confirm the agreement. 

For the $\SO(10)$ spinor $\mathbf{16}$, we have \begin{multline}
X_{\mathbf{16}}(x_i;w_i;m)=-\frac12 x_1^2 -\frac14 w_2 x_1 -\frac14(w_4-\frac14w_2^2)+\frac12 x_3\\ -m x_2 +(x_1+\frac12 w_2) m^2 + m^4\label{16}
\end{multline} The massless limit was determined in \cite{Aganagic:1997wp} and the massive case was found in \cite{Terashima:1998fx}.

For the $\SO(12)$ spinor $\mathbf{32s}$, we have \begin{multline}
X_{\mathbf{32s}}(x_i;w_i;m)=\frac{1}{256}(8 i x_2 + 8 x_1^2 + 4 w_2 x_1 + 4 w_4- w_2^2 )^2 \\
+m^2(i\tilde w_6 + w_6 - \frac14 w_2 w_4 + \frac 1{16} w_2^3+2w_2 x_1^2 -\frac 18 w_2^2 x_1 + \frac32 w_4 x_1 + 3 x_1^3 + 3i x_1 x_2 + \frac i2 w_2 x_2) \\
+4 i m^3 x_3 + m^4( 3i x_2 + \frac12 w_2 x_1 + \frac38 w_2^2-\frac12 w_4) + (2 x_1 + w_2)m^6+ m^8.\label{32}
\end{multline} The polynomial for the half-hypermultiplet is then \begin{equation}
X_{\frac12\mathbf{32s}}(x_i;w_i)=\frac{1}{16}(8 i x_2 + 8 x_1^2 + 4 w_2 x_1 + 4 w_4- w_2^2 ).
\end{equation} The polynomials for the conjugate spinor $\mathbf{32c}$ can be obtained by applying the outer automorphism $x_2\to -x_2$ and $\tilde w_6\to -\tilde w_6$:
 \begin{multline}
X_{\mathbf{32c}}(x_i;w_i;m)=\frac{1}{256}(-8 i x_2 + 8 x_1^2 + 4 w_2 x_1 + 4 w_4- w_2^2 )^2 \\
+m^2(-i\tilde w_6 + w_6 - \frac14 w_2 w_4 + \frac 1{16} w_2^3+2w_2 x_1^2 -\frac 18 w_2^2 x_1 + \frac32 w_4 x_1 + 3 x_1^3 - 3i x_1 x_2 - \frac i2 w_2 x_2) \\
+4 i m^3 x_3 + m^4( -3i x_2 + \frac12 w_2 x_1 + \frac38 w_2^2-\frac12 w_4) + (2 x_1 + w_2)m^6+ m^8.\label{32c}
\end{multline}

 The massless limit was first obtained in \cite{Aganagic:1997wp} and the massive case was found in \cite{Hashiba:1999ss}.
For both  $\SO(12)$ and $\SO(10)$ spinors, it should be possible to extend the analysis of \cite{Tachikawa:2010vg,Chacaltana:2011ze}  to find the realization via 6d $D_{6}$ and $D_5$ theory and compare the resulting curves. 

\subsection{$X_R$ for  $E_n$}

The ALE space for $E_6$ is given by \begin{equation}
W_{E_6}= x_1^4+ x_2^3+x_3^2+ w_2 x_1^2 x_2 + w_5 x_1 x_2 + w_6 x_1^2 + w_8 x_2 + w_9 x_1 + w_{12}.
\end{equation}  and the one for $E_7$ is 
\begin{multline}
W_{E_7}=  x_1^3{x_2} +x_2^3+x_3^2 \\
 + w_2 {x_2}^2 {x_1} + w_6 {x_2}^2  + w_8 {x_2} {x_1} + w_{10} {x_1}^2 + w_{12} {x_2} + w_{14} {x_1}+ w_{18}.
\end{multline}
The relation between $w_k$ and the Cartan $\phi$ is given e.g.~in Appendices of \cite{KatzMorrison}. Our normalization is slightly different from theirs: for $E_6$, our $w_k$ and their $e_k$ are related as \begin{equation}
w_2=e_2,\ 
w_5=-\frac{e_5}4,\ 
w_6=\frac{e_6}4,\ 
w_8=\frac{e_8}{16},\ 
w_9=-\frac{e_9}{16},\ 
w_{12}=\frac{e_{12}}{64}.
\end{equation} For $E_7$, the relations are \begin{equation}
w_2=-\frac{e_2}4,\ 
w_{6}=\frac{e_6}4,\ 
w_8=\frac{e_8}{16},\ 
w_{10}=\frac{e_{10}}{64},\ 
w_{12}=-\frac{e_{12}}{16},\ 
w_{14}=-\frac{e_{14}}{64},\ 
w_{18}=\frac{e_{18}}{64}.
\end{equation}

For $\mathbf{27}$ of $E_6$, the polynomial is given by \begin{multline}
X_{\mathbf{27}}(x_i;w_i;m)= -8(x_1^2-ix_3+\frac12w_6) -4 w_2 x_2 \\
+4mw_5+m^2(w_2^2-12 x_2)-8m^3 x_1 + 2m^4 w_2 + m^6.\label{27}
\end{multline} The massive case was determined in \cite{Terashima:1998iz}.
For $\mathbf{56}$ of $E_7$, we have \begin{multline}
X_{\mathbf{56}}(x_i;w_i;m)= x_2^2 + m^2(-6 x_1x_2 - 4 w_{10}) + 8im^3 x_3 \\
+m^4(-3x_1^2-6w_2 x_2-4w_8) + m^6(2w_2x_1 -10 x_2 -4 w_6) \\
+m^8(6x_1+w_2^2) + 2 w_2 m^{10}+m^{12} \label{56}
\end{multline} The polynomial for the half-hypermultiplet is then \begin{equation}
X_{\frac12\mathbf{56}}(x_i;w_i)=x_2.
\end{equation} The massless case was determined in \cite{Brodie:1997qg}. The massive case was determined in \cite{Hashiba:1999ss}.

\section{Seiberg-Witten curve from ALE geometry}\label{cg}


As explained, the Seiberg-Witten curve can be constructed from 
the $z$ dependent Casimirs $\tilde{w}$.
If we found that the geometry can be written as (\ref{tw}),
then 
by the redefinition
$x'_1=k(z)^{d_1} \tilde{x}'_1,
x'_2=k(z)^{d_2} \tilde{y}'_2, 
x'_3=k(z)^{d_3} \tilde{y}'_3$,
where $k(z)$ is an arbitrary function $z$ and $d_i$ is the dimension of
$x'_i$,
the geometry has the form of (\ref{tw}) again.
This can be considered as the ambiguity of the Seiberg-Witten form $\lambda$,
which we fixed, because
this is just the coordinate change.
Thus, in order to read off the $\tilde{w}$,
we should fix this ambiguity.
If we know the map between the type IIb description and 
6d (2,0) theory description, we can fix it.
However, for the cases we considered, we do not know it and
instead of it we will assume that
the Seiberg-Witten form satisfied the usual properties
which are related to the positivity of the effective coupling constant.

In order to do that, 
we now study the behavior of the $\lambda$ near the singularities.
If the solution $x(z)$ of $P_R(x,\tilde{w}_i)=0$ near $z=0$ 
scales as 
\beq
\left( \frac{x}{M} \right)^a \sim z^b,
\eeq
where $M$ is a dimension one constant and $a,b$ are 
positive co-prime numbers,
we introduce $\zeta^a=z$ which means $x/M \sim \zeta^b$.
Then, we find 
\beq
\lambda \sim a M \zeta^{b-1} d \zeta. 
\eeq
As usual, it will be required that 
allowed singularity is single pole and the residue is proportional 
to the mass of the theory.
This implies $b=0$ and $M$ is a linear combination of 
the mass parameters (or $b \geq 1$).
This means that
at $z=0$ and $z=\infty$,
$\tilde{w}_i$ should take finite value which does not depend
on $w_i$.

At another singularity $z=z_0$,
we can assume that the solution $x(z)$ behaves
\beq
\left( \frac{x}{M} \right)^a \sim \frac{1}{h^b},
\eeq
where $M$ is a dimension one constant and $a,b$ are 
positive co-prime numbers.
Here $h$ is the deviation from the singularity: 
$z/z_0-1=- h$ where $c$ is a constant.
We introduce $\zeta^a=h$ which means $x/M \sim \zeta^{-b}$.
Then, we find 
\beq
\lambda \sim  -M \zeta^{a-b-1} d \zeta. 
\eeq
Thus, we require that $a-b \geq 1$ or 
$a=b(=1)$. For the latter case, $M$ should be independent of $w_i$.
Another thing usually required is the condition
\beq
\frac{\partial \lambda}{\partial w_i} = {\rm a \, holomorphic \, form},
\label{cond1}
\eeq
which indeed follows from the above requirements.
If this is satisfied, then, the positive definiteness 
of the effective coupling constant is assured.

We have explicitly studied the several Seiberg-Witten geometries
and found that the ambiguity is indeed fixed uniquely 
by the requirements for the geometries we studied.
Note that it is highly non-trivial to find the $\tilde{w}_i$
which satisfies the requirements. Thus, 
the fact that we indeed found those $\tilde{w}_i$ is the 
evidences for the validity of the Seiberg-Witten geometry we proposed.

Finally, we will write down the explicit coordinate change 
for finding $\tilde{w}$ for some examples.
For the geometry (\ref{bosh}) of the full $\mathbf{20}$ of $SU(6)$,
the correct coordinate change is
\beqa
x'_1 &=& \frac{1}{\sqrt{h}} x_1 \CR 
x'_{{2}} &=& \frac{1}{h} \left(
x_{{2}}+4\,i \,{m}^{3} z  \right) \CR
x'_{{3}} &=& \frac{1}{\sqrt{h}} \left(
x_{{3}}-2\,i \frac{1}{h}
\, \left( 3\,{m}^{2}x_{{1}}+w_{{2}}x_{{1}}+2
\,{x_{{1}}}^{3}+w_{{3}} \right)  z \right),
\eeqa
where 
\beq
h=1+4 z,
\eeq
the Seiberg-Witten geometry becomes
\beq
h^{2} \, W_{A_5}(x'_1,x'_2,x'_3;\tilde{w}) = 0,
\eeq
where $\tilde{w}_i=\tilde{w}_i(z)$ satisfies the above conditions
and the correct classical limit $\tilde{w}_i \rightarrow w_i$ with 
$z \rightarrow 0$.
Of course, we can change $z \rightarrow -z/4$ in order to 
set the singularity at $z=1$.
There are artificial cuts at the singular points $h=0$
and $h=\infty$, but, we find that the $\tilde{w}_i$ is uniquely defined
if we think $w_i$ transforms to $(-1)^i w_i$ when we go around 
the singular points.
Thus, despite the introduction of the artificial cut at the singular points $h=0$
and $h=\infty$,
we believe that
that these $\tilde{w}_i$ give the correct curve.

For the geometry of the $\mathbf{56}$ of $E_7$,
the correct coordinate change is
\beq
x'_1=\frac{x_1- m^4 z}{h},
\;\; x'_2=\frac{x_2}{h}, \;\; 
x'_3=x_3 +4 i m^3 x_2 z,
\eeq
where $h=z+1$,
the geometry becomes
\beq
h^4
W_{E_7}(x'_1,x'_2,x'_3;\tilde{w}) = 0,
\eeq
where $\tilde{w}_i$ satisfies the conditions.

The last example is the the geometry of the $\mathbf{27}$ of $E_6$:
\begin{equation}
  \label{swgeometrye6b}
    z X_{{\bf 27}}(x_i;w_i;m)^{2} 
 + W_{E_6}(x_1,x_2,x_3;w) = 0,
\end{equation}
where we have set the all mass parameter are same.
Using the coordinates:
\beqa
x'_1 &=& \frac{x_1+32 m^3 z}{h} \CR 
x'_2 &=& \frac{h x_2+\frac{16}{3} z (w_2+3 m^3)^2 }{h^2}, \CR
x'_3 &=& \frac{h x_3+ 8 i z (X_{E_6}(x_1,x_2,x_3=0))
}{h^2},  
\eeqa
where 
\beq
h=1-64 z,
\eeq
the SW geometry becomes
\beq
h^3 \, W_{E_6}(x'_i;\tilde{w}) = 0,
\eeq
where $\tilde{w}_i=\tilde{w}_i(z)$ satisfy following conditions:
\begin{itemize}
 \item correct classical limit
\beq
\tilde{w}_i \rightarrow w_i, \; as \; z  \rightarrow 0,
\eeq
 \item finite at $z=\infty$. Indeed, at $z=\infty$ we find\footnote{
If two mass parameters are different, say $m_1,m_2$, it is still finite,
but complicated. For following two examples, we found some simple results. 
For $m_1=-m_2=m$,
we find $W_{E_6}(x'_1,x'_2,x'_3;\tilde{w}) =
{x'_1}^4+{x'_2}^3+{x'_3}^2 -3 m^2 {x'_1}^2 x'_2+m^6 {x'_1}^2$.
For $m_2=0$, the characteristic polynomial is $x^9(x^2-4 m^2)^3 (x^2-m^2)^6$.}
\beq
 W_{E_6}(x'_1,x'_2,x'_3;\tilde{w}) =
{x'_1}^4+{x'_2}^3+{x'_3}^2-3m^2 {x'_1}^2 x'_2+3 m^5 x'_1 x'_2 -\frac{3}{2} m^6
       {x'_1}^2-\frac{3}{4} m^8 x'_2 +m^9 x'_1 -\frac{3}{16} m^{12}.
\eeq
 \item at the singular point $h=0$, 
the leading behavior is
\beqa
 W_{E_6}(x'_1,x'_2,x'_3;\tilde{w}) &\sim &
{x'_1}^4+{x'_2}^3+{x'_3}^2 
-\frac{3m^2+w_2}{h} {x'_1}^2 x'_2
 -\frac{1}{12} \frac{(3 m^2+w_2)^3}{h^3} {x'_1}^2 \CR && 
-\frac{1}{48} \frac{(3 m^2+w_2)^4}{h^4} x'_2  
+\frac{1}{864} \frac{(3 m^{2}+w_2)^6}{h^6}.
\eeqa
The characteristic polynomial $P_{\mathbf{27}}$ with the $\tilde{w}$ becomes
$x^{15} (x^2-(6 m^2+2 w_2)/h)^2$ near $h=0$.
\end{itemize}

\bibliographystyle{ytphys}
\small\baselineskip=.9\baselineskip
\bibliography{ref}
\end{document}